\documentclass{rspublic}
\pdfoutput=1

\usepackage{graphicx}

\begin{document}

\title[Cosmic String Networks]{Small Scale Behavior of Cosmic String Networks}

\author[J. Polchinski]{Joseph Polchinski}

\affiliation{Kavli Institute for Theoretical Physics\\ University of California, Santa Barbara, CA 93106-4030, USA}

\label{firstpage}

\maketitle

\begin{abstract}{Cosmic String Networks, Cosmology}
The evolution of cosmic string networks is an interesting dynamical problem.  The equations governing these networks are classical and fully specified, but the length scale at which cosmic string loops form has been uncertain to tens of orders of magnitude.  Numerical simulations are limited in the range of length and time scales they can reach, while analytic methods have been limited by the nonlinearities of the problem.  We describe a recent analytic scaling model developed in collaboration with Jorge Rocha and Florian Dubath which, together with recent simulations, appears to resolve this question.	

Presented at the Royal Society Discussion Meeting ``Cosmology Meets Condensed Matter'', January 28-29, 2008.
\end{abstract}

\section{Introduction}

The formation of string networks in phase transitions is a well-known connection between cosmology and condensed matter physics.  I would like to speak about a further connection which is not so widely 
appreciated, that is, the evolution of these networks {\it after} they form (Albrecht \& Turok, 1989; Allen \& Shellard 1990; Bennett \& Bouchet, 1990).

Figure~1 shows a simulation of what such a string network might look like today.  
\begin{figure}
\begin{center}
\includegraphics[scale=0.25]{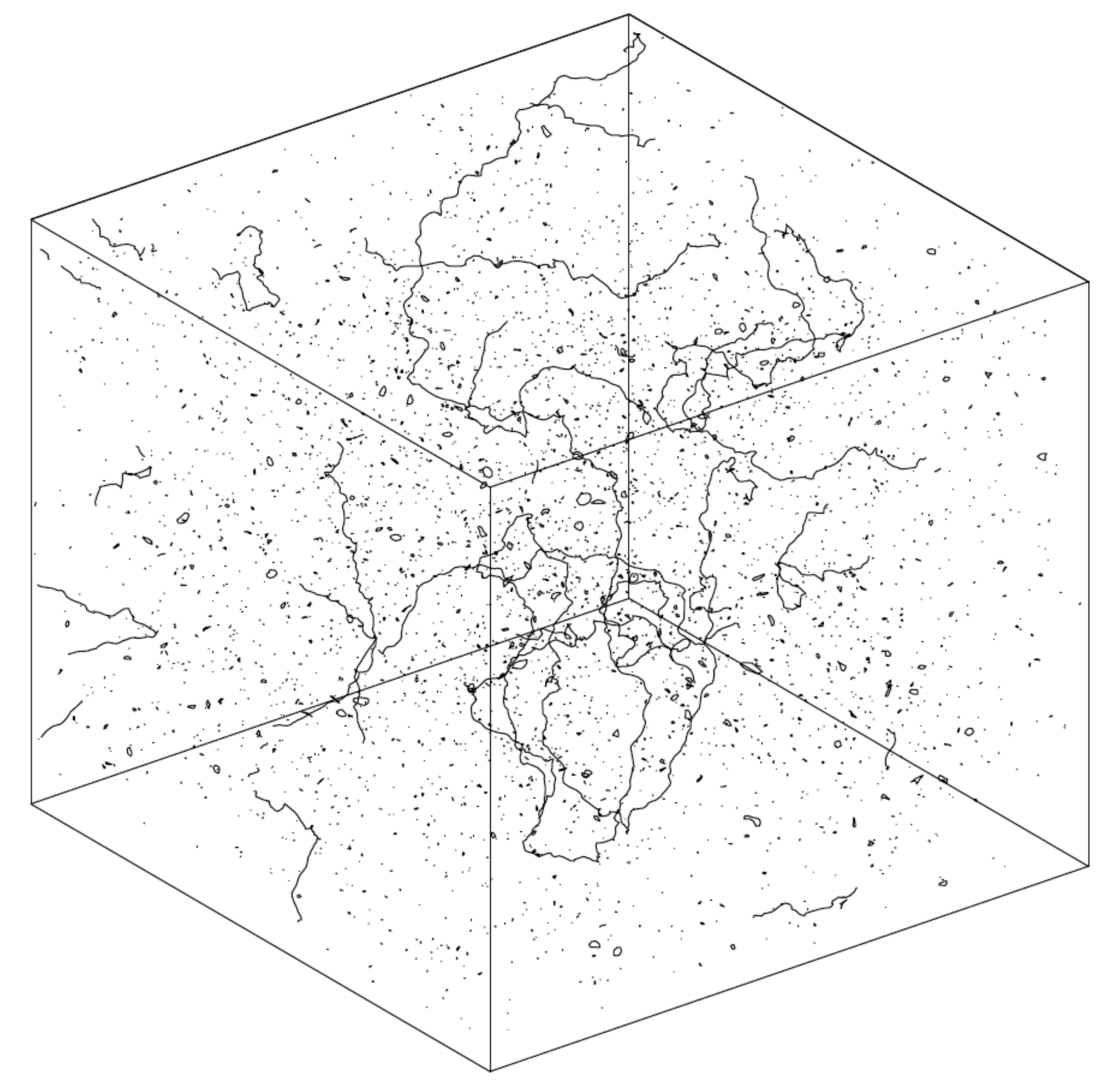}
\end{center}
\caption{Simulation of cosmic string networks, from Allen \& Shellard (1990).  The side of the cube is of order the Friedman-Robertson-Walker (FRW) time.} 
\end{figure}
Qualitatively it is similar to the network at formation, with some of the string in the form of infinite random walks and some in the form of loops of various sizes.  However, to understand the signatures it is essential to be quantitative.  For example, what is the mean density of string, and what is the distribution of loop sizes?
It has been understood since the very early work of Kibble (1976, 1980) that most or all of the properties of the network are actually independent of the details of formation: the statistical properties of the network are given by an attractor solution.

The quantitative nature of this attractor solution is still very uncertain.  There is reasonable agreement (though not unanimity) about the total density in long strings.  However, estimates of the typical size of loops vary over tens of orders of magnitude, from nearly the size of the physical universe down to solar system scales and even down to microphysical scales.  Remarkably, this true even when the equations governing the network are completely specified: it reflects our inability to solve these equations, either by numerical or analytic means.   The closest analogy I can think of is turbulence.

\section{Strings and signatures}

Let me first briefly review what kind of strings we are talking about, and how they might be observed.
For a more extended review see Polchinski (2008).  The classic cosmic strings are gauge theory solitons, magnetic flux tubes, but the fundamental strings of superstring theory might also appear as cosmic strings.  Other possibilities include the dual D-strings, as well as a variety of higher dimensional branes partially wrapped so that that only one dimension is extended; one could also have electric flux tubes in the low energy gauge theory.  

Since the strings are of cosmic length and microscopic thickness, they will all be described to first approximation by the Nambu action.  However, there are a number of macroscopic properties and parameters that might distinguish them.  The first is their tension $\mu$.  Here it is useful to work with the dimensionless combination $G\mu$, which is a measure of the typical metric perturbation produced by the string.  Another distinguishing property is what happens when two strings collide, since this is sensitive to their internal structure: do they pass through one another, or do they reconnect?  The probability of reconnection is denoted $P$; to be precise, it is a function of the collision angle and velocity, but one usually takes a mean value.  Other possible distinguishing properties would be the string having additional long-ranged interactions beside gravity (e.g. axionic, or gauge); the presence of extra low energy degrees of freedom, beyond the collective coordinates for its three-dimensional motion; the existence of more than one kind of string; and, the existence of junctions where three or more strings meet.

If cosmic strings are ever seen, the first task will be to determine these macroscopic properties, and then to try to deduce the underlying microscopic structure.   There is one notable situation in which there is a simple correspondence: if the only two models were the gauge theory solitons and the fundamental strings, then $P$ would distinguish them: it is 1 for the gauge theory strings, where reconnection is a classical process (Shellard, 1987), and $O(g_{\rm s}^2)$ for fundamental strings, where it is a quantum process (Polchinski, 1988).  

In order to determine the macroscopic properties from observations, a precise quantitative understanding of the networks will be needed.  In fact,  the uncertainties are are large even for the simplest `vanilla' networks, with one kind of string, no junctions or extra long-range interactions or degrees of freedom, and $P=1$.  We will focus on this simplest case.

For vanilla networks the signatures are all gravitational: the effect on the cosmic microwave background (CMB), gravitational lensing, and gravitational radiation.  The perturbations of the CMB come primarily from the long strings, for which the uncertainties are not so large but could still be improved.  Long strings will lens, and so will loops if they are large enough.  Gravitational radiation comes predominantly from the loops, and the intensity and frequency depend crucially on the highly uncertain loop size distribution.  This radiation is likely to be the strongest signature, and so it is essential to resolve this question.

\section{Network evolution}

In this section we describe the processes governing the evolution of vanilla string networks.  For further details and references see Polchinski (2008), and also the classic reviews by Vilenkin \& Shellard (1994) and by Hindmarsh \& Kibble (1995).

\subsection{Formation}

Any gauge theory with a broken $U(1)$ symmetry has a string soliton solution.  Moreover, such strings will actually form in any phase transition in which a $U(1)$ symmetry {\it becomes} broken (Kibble, 1976), and a finite fraction of the string will be in infinite random walks.  These infinite strings are the seeds of the later network, because any loops that form in the transition decay rapidly.
Subsequently, Sarangi \& Tye (2002) made the remarkable observation that D-strings would be produced by D-brane annihilation at the end of brane inflation, since the low energy dynamics is again described by $U(1)$ symmetry breaking.  In fact, regardless of the microscopic structure of the strings, there is some dual version of the Kibble process, such that if the strings can exist only after some phase transition, they will actually form during that transition.  

\subsection{Stability}
We must assume that the processes of breakage and string confinement are either absent or sufficiently suppressed so as to be negligible on cosmological time scales.

\subsection{Strings in expanding spacetime}

Cosmic strings are long compared to their thickness, and so can be treated as idealized one-dimensional objects.  For vanilla strings, the relevant action is just the Nambu action in the FRW metric
\begin{eqnarray}
\rd s^2 &=& -\rd t^2 + a(t)^2 \rd {\bf x} \cdot \rd {\bf x} \nonumber\\
&=& a(\tau)^2 (-\rd \tau^2 + \rd {\bf x} \cdot \rd {\bf x})\ .
\end{eqnarray}
The equation of motion governing the evolution of a cosmic string is
\begin{equation}
\ddot{\bf x} + 2\,\frac{\dot{a}}{a}\,(1-\dot{\bf x}^2)\, \dot{\bf x} = \frac{1}{\epsilon} \left( \frac{{\bf x}'}{\epsilon} \right)' \ .
\label{EOM}
\end{equation}
Here $\epsilon$ is given by
\begin{equation}
\epsilon \equiv \left( \frac{{\bf x}'^2}{1-\dot{\bf x}^2} \right)^{1/2}.
\label{epsilon}
\end{equation}
These equations hold in the transverse gauge, where $\dot{\bf x}\cdot{\bf x}'=0$. Dots and primes refer to derivatives with respect to the conformal time $\tau$ and the spatial parameter $\sigma$ along the string, respectively. The evolution of the parameter $\epsilon$ follows from equation (\ref{EOM}),
\begin{equation}
\frac{\dot{\epsilon}}{\epsilon} = -2\,\frac{\dot{a}}{a} \, \dot{\bf x}^2.
\label{epsev}
\end{equation}

The equation of motion implies that on scales long compared to the horizon length, which is of the order of the FRW time $t$, the string is frozen in the comoving coordinates $\bf x$, and so just expands with the universe.  When a scale comes inside the horizon, the modes begin to oscillate; the energy begins to redshift away, and the string straightens.

From the second derivative terms it follows that signals on the string propagate to the right and left with $d\sigma = \pm d\tau/\epsilon$.  Thus the structure on a short piece of string at a given time is a superposition of left- and right-moving segments.   In an expanding spacetime the left- and right-moving waves interact --- they are not free as in flat spacetime.  It is very useful to discuss the motion of the string in terms of left- and right-moving unit vectors 
\begin{equation}
\bf{p}_\pm \equiv \dot{\bf{x}} \pm \frac{1}{\epsilon}\bf{x}'\ .
\end{equation}
The equation of motion~(\ref{EOM}) in terms of these is
\begin{equation}
\dot{\bf p}_\pm \mp \frac{1}{\epsilon} {\bf p}'_\pm = - \frac{\dot{a}}{a} \left[ {\bf p}_\mp - ({\bf p}_+ \cdot {\bf p}_-)\, {\bf p}_\pm \right],
\label{EOM2}
\end{equation}
so that in the limit that we can ignore the time derivative of the scale factor we recover independent left- and right-moving waves.

\subsection{Long string reconnection}

When two straight strings reconnect, the resulting strings have a kink, which separates into left- and right-moving kinks.  Under the flat spacetime Nambu equations, the kinks would move indefinitely in the two directions.  The curved spacetime equations imply that the kinks persist but their angle slowly decreases.  The decrease is sufficiently slow that the kinkiness is potentially very large: if one considers the total root-mean-square kink angle in a length of long string (which is what would be relevant if the kink directions were uncorrelated), one finds that it diverges, so there is the potential for having structure on very small scales in the network.

\subsection{Long string gravitational radiation}

The oscillations of the long string will emit gravitational radiation, and so are damped below a certain length scale.  A naive estimate would lead to the same scale that we will find below for the loop decays, something of order the dimensionless string tension $G\mu$ times the FRW time $t$.  However, things are more subtle for long strings.  Emission of gravitational radiation requires that a right- and left-moving mode meet.  There is a nonlinear suppression when these have very different wavelengths, because the short-wavelength modes essentially perceives a straight string, so that the actual damping scale is $O(t)$ times a larger power of $G\mu$. 

\subsection{Loop formation}

When a long string intercommutes with itself, a loop breaks off.  The loop may then self-intercommute and fragment into smaller loops.  The fragmentation happens rapidly, within one period, so we are interested in the size distribution of the ultimate non-self-intersecting loops.
Of course a loop may collide with a long string (or each other) and reattach; for loops close to the FRW scale this is likely, but smaller loops have a high probability never to rejoin.

\subsection{Loop decay}

Dimensionally, a loop of length $l$ emits gravitational radiation power of order $G \mu^2 $.  Since its total energy is of order $\mu l$, we conclude that the lifetime is
\begin{equation}
t(l) = l/\Gamma G \mu \ ,
\end{equation}
where the numerical factor $\Gamma$ is found to be of order 50 for typical loops.  In other words, loops smaller than $\Gamma G \mu t$ will decay in a Hubble time.

\section{Scaling, loop size, and the condensed matter connection}

If the network simply followed the expansion of the universe, all distances would grow as the scale factor~$a(t)$, which goes as $t^{1/2}$ during the radiation-dominated era and as $t^{2/3}$ during the matter-dominated area.  However, the net effect of the evolution processes is to remove string from the network, so that the typical separation grows more rapidly.  The horizon length (which is the same as the FRW time and as the Hubble time, up to numerical factors) is the maximum distance over which causal processes could have acted, and so is the fastest possible growth rate for the string separation.  In fact, analytic arguments and simulations show that that this maximum is attained.  

The {\it scaling hypothesis} states that the statistical properties of the network approach are constant when lengths are scaled to $t$.  If the long string density exceeds the attractor value, the rate of collisions increases, resulting in more kinks, more loop formation, and less long string; if it is smaller, than the opposite occurs.  This means that the initial conditions from the formation of the network, in particular the initial density of strings, are washed out in time.

The scaling hypothesis implies that the typical size at which loops form is a multiple of $t$.  However, this leaves open the question as to whether the ratio is a pure number, or depends on the (small) dimensionless parameter $G\mu$.  The physical question is whether the smoothing of the network by gravitational radiation (which takes place on a scale parametrically smaller than the horizon) is important in determining the loop size, or whether the nonlinear dynamics of the network at scales near the horizon is sufficient.  For the interesting values of $G\mu$, this represents an uncertainty of twenty orders of magnitude.

This remains a contentious subject even after twenty-five years of study.  If the horizon scale had been the only relevant scale in the problem, simulations would have readily determined the network properties.  Instead, it is found that interesting physics takes place at the limits of spatial resolution and simulation time, and so one must attempt to extrapolate: is one seeing scales that are pure numbers, perhaps small numbers, times the horizon scale, or is new physics (and a new parameter, $G\mu$) needed to attain scaling?

Unfortunately, analog systems seem difficult to employ here.  The bare minimum physics would be the expansion of the universe and the reconnection of colliding strings.  However, it is important that the dynamics is Lorentz invariant and nearly dissipationless; losing either of these properties would likely change the network behavior substantially.  Also, it is important to reach large dynamic ratios of length and time.  It would seem difficult to find an analog system that does better than the numerical simulations.  Rather, I think that the useful condensed matter parallel is one of theoretical approach.

Analytic methods are difficult because of the nonlinearities of the problem.  One can write down the evolution equations for an ensemble of strings, but before long one must begin to approximate, and in the end most analytic treatments reduce to somewhat coarse models in terms of a few parameters.  On the other hand, one might have expected that the nature of the problem, the existence of a large ratio of scales, would lend itself to something like a renormalization group treatment.  Roughly speaking, one would use the simulations to treat the strongly nonlinear behavior at the Hubble scale, and then evolve down to shorter scales analytically.

This was the motivation for the work that will be described in the next section.  Thus far, things are not as systematic as for the renormalization group.  The difficulty is largely that for the string network the flow is not from short distances to long but the other way around.  The Hubble length, against which all things are scaled, increases more rapidly than the comoving length (during the matter and radiation eras), so comoving structures move to smaller effective scales over time.  In this respect the problem is like turbulence, and has been similarly frustrating.

Nevertheless, it has been possible to proceed by making approximations that are not precisely controlled but perhaps well-motivated.  In particular, this has allowed us to understand why structures appear at scales far below the horizon length.  The simulations are still essential, but the analytic model enables us to distinguish real effects from transient ones, and to extrapolate beyond the scales that can be simulated.  Combined with recent simulations that use a trick to get beyond the limitation on expansion times, it may be that a consistent and fairly quantitative picture is emerging.

\section{A model of short distance structure}

The model we describe is developed in Polchinski \& Rocha (2006, 2007) and Dubath {\it et al.} (2007), to which the reader is referred for more detail.  The philosophy is that the strongly nonlinear dynamics at the horizon scale (and possibly at some other scales) must be left to the simulations, but that the subsequent evolution over wide ranges of scale can be treated analytically.

\subsection{The long-string two-point function}

We first consider the small-scale structure on the long strings.  To do this we focus on the evolution of a short left- or right-moving segment on a long string.  This will potentially involve the following:
\begin{enumerate}
\item
Evolution according to the FRW-Nambu equations~(\ref{EOM2}).
\item
Long string intercommutation.
\item
Incorporation into a loop larger than the segment.
\item
Emission of a loop comparable to or smaller than the segment.
\item
Smoothing via gravitational radiation.
\end{enumerate}
The probability of the second of these is proportional to the length of the segment, and so can be systematically neglected for a short segment.  The third process is governed by dynamics on a longer distance scale, and so does not depend directly on the configuration of the segment: it will not change the statistical distribution for configurations of small segments.  

In saying this we are neglecting the correlation between the structure at short distance and the structure at long distance.  This is not a fully controlled approximation, but should become progressively better at shorter scales.  One could attempt to make an improved model in which this correlation is parametrized, but this will not be necessary for our purposes.  Ultimately one would like to write down the exact equations for an ensemble of string networks, and then approximate systematically, but this is apt to be very difficult; it is more practical to first figure out what is going on.  

The small loop production would be absent if the only scale were the Hubble length.  We will see that if we assume this and proceed, it will not be self-consistent.  However, we will also find that the production of small loops is primarily controlled by the long-distance structure rather than the short distance structure, so we seem to get the right answer anyway.  Finally, we will assume that, while we are below the Hubble scale, we are at large enough scales to ignore gravitational radiation.

Thus we need only to solve for the motion of a string in an expanding universe.  Equations~(\ref{EOM2}) are not solvable in closed form, but they simplify for a short segment.  Separate the configuration into a mean direction of the segment and a deviation,
\begin{equation}
{\bf p}_\pm(\tau,\sigma) = {\bf P}_\pm(\tau) + {\bf w}_\pm(\tau,\sigma)
-\frac{1}{2}  {\bf P}_\pm(\tau) {w}^2_\pm(\tau,\sigma) + \ldots\ ,
\label{ppw}
\end{equation}
where $P_{\pm}^2 = 1$ and ${\bf P}_\pm \cdot {\bf w}_\pm = 0$.  Expanding in powers of ${\bf w}_\pm(\tau,\sigma)$ gives
\begin{eqnarray}
\dot{\bf P}_+ &=& - \frac{\dot{a}}{a} \left[ {\bf P}_- - ({\bf P}_+ \cdot {\bf P}_-)\, {\bf P}_+ \right]\ ,
\\
\dot{\bf w}_+ - \frac{1}{\epsilon} {\bf w}'_+ 
&=& - ({\bf w}_+ \cdot \dot{\bf P}_+)\, {\bf P}_+ +  \frac{\dot{a}}{a} ({\bf P}_+ \cdot {\bf P}_-)\, {\bf w}_+\ .
\label{EOM3}
\end{eqnarray}
In Eq.~(\ref{EOM3}) we have dropped a term $\frac{\dot{a}}{a} ({\bf P}_+ \cdot {\bf w}_-)\, {\bf P}_+$, because it averages to zero (with corrections suppressed by the length of the segment) as the left- and right-moving waves sweep past each other.  The first term on the right of equation~(\ref{EOM3}) is just a precession, which keeps ${\bf w}_+$ perpendicular to ${\bf P}_+$; if we work with parallel transported axes we can ignore it.  In the second term, the factor of $\dot a/a$ means that the variation is significant only over a Hubble time.  We then replace ${\bf P}_+ \cdot {\bf P}_-$ with its ensemble-averaged value. This is simply $2\bar v^2 - 1$ with $\bar v^2$ the ensemble average, which we obtain from the simulations, $\sim 0.41$ in the radiation era and $0.35$ in the matter era.  Again, this is not a fully controlled approximation, but should become progressively better for shorter segments.

Along a left-moving characteristic we then have ${\bf w}_+ \propto a^{2\bar v^2 -1} \propto t^{r(2\bar v^2 -1)}$, and similarly for ${\bf w}_-$ along a right-moving characteristic.  Here we take $a = t^r$, so that $\tau \propto t^{1-r}$.  Averaging over a translation-invariant ensemble of segments gives
\begin{equation}
\left\langle [{\bf w}_+ (\sigma,\tau) - {\bf w}_+(\sigma',\tau)]^2 \right\rangle
= t^{-2r(1 - 2\bar v^2 )} f(\sigma-\sigma')\ .
\label{ww}
\end{equation}
The physical length $l$ of a segment of coordinate length $\delta \sigma$ is 
\begin{equation}
l = a(t) \epsilon(t) \delta \sigma \sim a(t)^{1 - 2\bar v^2} \delta \sigma \ , \label{physl}
\end{equation}
 where we have averaged in the $\epsilon$ equation of motion~(\ref{epsev}).
  As time goes on, the physical length of the segment increases as $a^{1 - 2\bar v^2}$, but this is much slower than the growth of the Hubble length $O(t)$, so the segment `propagates' to shorter and shorter scales.  Correspondingly, as we evolve back, there will be a point at which its length approaches $t$ and our approximations break down.  At this point, we have to let the simulations deal with the horizon-scale dynamics, providing an initial condition that determine the constant $t$.  That is, the correlator~(\ref{ww}) takes some value when $a^{1 - 2\bar v^2} (\sigma - \sigma') = Ct$, where the constant $C$ defines the matching scale.  This constant will be independent of time if the Hubble-length dynamics scales, so we can conclude that
\begin{equation}
f(\sigma - \sigma') \propto  |\sigma-\sigma'|^{2 \chi}\ ,\quad
\chi = \frac{ r (1 - 2\bar v^2 ) }{ 1 - r(1 - 2\bar v^2)}
\end{equation}
to cancel the
 time-dependence of $C$ at the matching point. 
In all,
\begin{equation}
\left\langle [{\bf w}_+ (\sigma,\tau) - {\bf w}_+(\sigma',\tau)]^2 \right\rangle
= 2{\cal A} (l/t)^{2\chi} \ .
\label{fluct}
\end{equation}
The parameters $\bar v^2$ and ${\cal A}$ must be taken from simulations, but the remaining functional form is determined.

This two-point function implies that the fractal dimension of the string approaches 1 a short distance (Martins \& Shellard, 2006), but slowly, as $1 + O([l/t]^{2\chi})$.   This power law behavior persists down to the scale where gravitational smoothing becomes important, $l_{\rm GW} \sim \Gamma {\cal A} (G\mu)^{1 + 2\chi_{\rm r}} t$  (Polchinski \& Rocha, 2007).

\subsection{Loop production}

We now take account of the production of small loops.  Since this takes place on scales small compared to the Hubble scale, we can go to locally flat null coordinates $u,v$, where ${\bf p}_+(u) = \partial_u {\bf x}$ and ${\bf p}_-(v) = \partial_v {\bf x}$, with $u + v = t$. 
A loop forms whenever a long string passes through itself, meaning that  $ {\bf x}(u,v+l) =  {\bf x}(u+l,v)$ for two points on the string.  Defining
\begin{equation}
{\bf L}_+(u,l) = \int_u^{u+l} du \, {\bf p}_+(u)\ ,\quad
{\bf L}_-(v,l) = \int_v^{v+l} dv \, {\bf p}_-(v)\ ,
\end{equation}
this means that ${\bf L}_+(u,l) = {\bf L}_-(v,l)$ for some $u$, $v$, and $l$.  The rate of loop formation, per unit volume in $u, v, l$, is then
\begin{equation}
\langle \det J\, \delta^3( {\bf L}_+(u,l) - {\bf L}_-(v,l) ) \rangle\ ,\quad J = \frac{\partial^3 ( {\bf L}_+(u,l) - {\bf L}_-(v,l) )}{\partial u\, \partial v\, \partial l}\ .
\end{equation}
Now, the components of ${\bf L_{\pm}}$ are each proportional to the length $l$, so the $\delta$-function implies a factor of $l^{-3}$.  The rate would scale as $l^{-3}$ if the correlator of the $\bf w$'s were scale invariant.  To work more carefully, separate ${\bf p}_\pm$ as in Eq.~(\ref{ppw}), with the unit vectors ${\bf P}_{\pm}$ proportional to ${\bf L}_{\pm}$.  Then for small ${\bf w}_\pm$,
\begin{equation}
\delta^3( {\bf L}_+(u,l) - {\bf L}_-(v,l) )  = l^{-2} \delta^2({\bf P}_+ -{\bf P}_-) \delta(L_+(u,l) - L_-(v,l)) \ .
\end{equation}
The magnitudes $L_\pm$ are both equal to $l(1 - O(l^{2\chi}))$, so the whole is of order $l^{-3-2\chi}$.  The columns of $J$ are of order $l^\chi, l^\chi, l^{2\chi}$ (the last from the components parallel to ${\bf P}_\pm$), so the determinant is of order $l^{4\chi}$ giving $l^{-3 + 2\chi}$ for the rate.

The total rate of string length going into loops is weighted by an additional factor of $l$,
\begin{equation}
\int dl\, l^{-2 + 2\chi}\ . \label{loopdiv}
\end{equation}
This diverges at the lower end for $\chi \leq \frac{1}{2}$, a crucial result.  This large production of small loops seems surprising at first, because the string is becoming smooth in the sense that the fractal dimension approaches 1.  What we have found above is that the production of small loops is controlled by the rate at which the fractal dimension approaches~1, and in both the matter and radiation eras the approach is slower than the critical value~$\chi = \frac{1}{2}$.

The total rate at which long string breaks off to form loops is fixed by energy conservation, so the divergence~(\ref{loopdiv}) must cut off in some way.  Physically, the loop production at a given point must stop when the probability of a given point breaking off reaches unity.   In Dubath {\it et al.} (2007) it is shown by separating the curves ${\bf p}_\pm$ into a classical long-distance piece and a random short distance piece (figure~2) that the loop production occurs near large scale cusps - that is, intersections of the classical ${\bf p}_\pm$ curves: for a small loop, equality of ${\bf L}_+$ and ${\bf L}_-$ occurs when the vectors ${\bf p}_{\pm}$ are nearly parallel.  This is why the production of small loops is determined by the long distance structure, as claimed earlier.  Further, it is found that the loop production distribution $l^{-2 + 2\chi}$ persists down to the gravitational radiation cutoff. 
\begin{figure}
\begin{center}
\includegraphics[width=14pc]{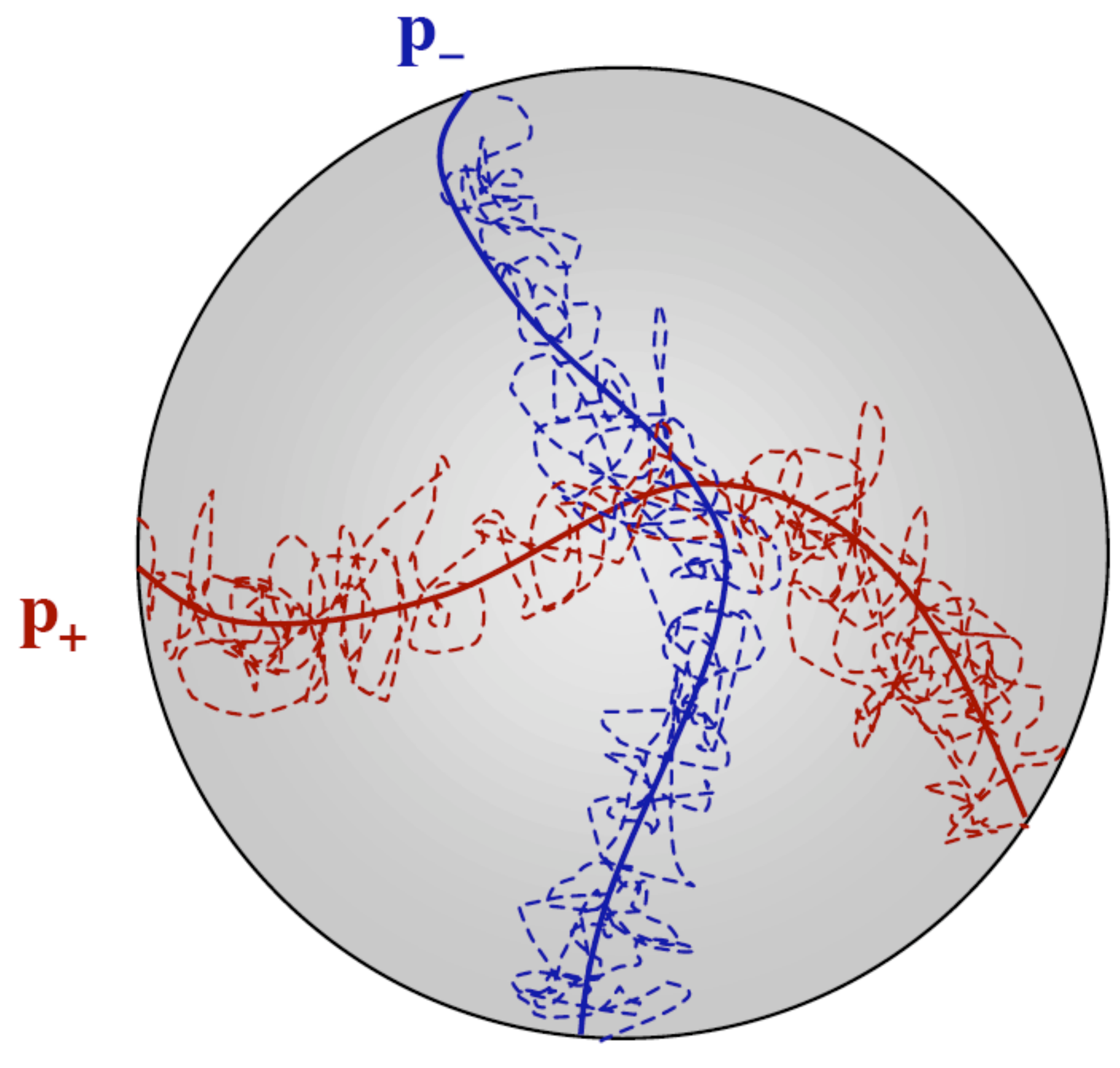}
\end{center}
\caption{The functions ${\bf p}_\pm$ written as a fixed long-distance piece plus a random short-distance part.  Production of small loops takes place near the cusp of the long-distance part.  Note that the figure is showing the tangent vector to the string, rather than the string itself.  The latter has a fractal dimension that approaches 1 at short distance, but the tangent vector path has a large fractal dimension as indicated.}
\end{figure}

The most recent simulations (Olum \& Vanchurin, 2007) show two peaks in the loop production.  One is at a scale of order $0.1 t$, while the other remains near the short-distance cutoff.  The authors of that work interpret the small peak as a transient, but it appears to match the analysis above: the exponent $-2 + 2\chi$ agrees quantitatively with the simulations, and also with those of Ringeval {\it et al.}(2007).  Thus we argue that these are a robust physical feature.  We cannot see the large loops in our model, which captures only the short-distance structure, but it is inevitable that some of these will arise from chance self-intersections of long strings.  Our picture does explain why these large loops can survive fragmentation:
one would expect from figure~2 that the regions near cusps will fragment extensively, but non-cusp regions of the long loop will remain.

Thus it appears that we may be approaching an accord on the question of loop sizes.  Some fraction of the string, perhaps a small fraction, goes into the largest scale under discussion, near the horizon.  The rest goes into loops at the gravitational radiation scale.\footnote{
A recent paper (Vanchurin, 2007) asserts that the small loop size does not depend on the gravitational radiation.  This paper contains several calculational errors that invalidate its conclusions. In particular, its quantity $\lambda_3(\Delta)$ is actually divergent, as follows immediately from the definition and the fact that $b < 2$ (note that the sign of $c$ must also be corrected).  This divergence has a simple origin.  At any kink, ${\bf a''}$ will have a delta-function contribution, so the RMS value will necessarily diverge.  By omitting this divergence, one essentially ignores the kinks, which will of course give an incorrect loop distribution. 

 In the real network, the kinks will be smoothed by gravitational radiation, so the actual value of $\lambda_3(\Delta)$ depends inversely on the gravitational radiation scale.  We believe that the methods of this paper are coarser than ours and cannot give as complete an account of loop formation, but it is notable that when the arithmetic error noted above is corrected the conclusions agree with ours: gravitational radiation enters in determining the small loop size.}
The simulations played an essential role in first revealing this double-peaked distribution, while the model plays a key role in confirming that the small loops are real, and in showing that gravitational radiation sets their actual size.

The divergence~(\ref{loopdiv}) explains why there is not just a single scale in the string network: the nonlinearities of string evolution have a remarkable ability to transfer energy from cosmic scales down to much shorter wavelengths.  In this respect the cusps are like shock waves.  Indeed, in shock waves the low energy field equations do break down, and shorter scale physics enters.  This has also been suggested for string networks (Vincent {\it et al.}, 1997).  We believe that smoothing due to gravitational radiation prevents this, but cosmic string networks have been a source of many surprises, and perhaps more are in store.

Finally, we emphasize the important lesson, that a precise understanding of cosmic string networks will depend on combining numerical and analytic calculations in the most effective way.

\section*{Acknowledgments}
I thank Jorge Rocha and Florian Dubath for collaborations, and the organizers and participants of this Royal Society Discussion Meeting for a stimulating event.  This work was  supported by NSF grants PHY05-51164 and PHY04-56556.

\end{document}